\documentclass[pdflatex,sn-mathphys-num]{sn-jnl}

\usepackage{graphicx}%
\usepackage{multirow}%
\usepackage{amsmath,amssymb,amsfonts}%
\usepackage{amsthm}%
\usepackage{mathrsfs}%
\usepackage[title]{appendix}%
\usepackage{xcolor}%
\usepackage{textcomp}%
\usepackage{manyfoot}%
\usepackage{booktabs}%
\usepackage{algorithm}%
\usepackage{algorithmicx}%
\usepackage{algpseudocode}%
\usepackage{listings}%

\theoremstyle{thmstyleone}%
\theoremstyle{thmstyletwo}%

\theoremstyle{thmstylethree}%

\newcommand{\msc}[1]{\normalfont\textsc{#1}}

\def\rv{\mathbf{r}}
\def\kth{k_\theta}

\def\vF{\mathrm{v}_{\msc{f}}}
\def\fbz{$1^{\mathrm{st}}$BZ}

\def\wM{\omega_{\msc{ma}}}
\def\thM{\theta_{\msc{ma}}}

\def\gaM{\gamma_{\msc{ma}}}

\def\GKMGt{\tilde{\Gamma}\tilde{K}\tilde{M}\tilde{\Gamma}}

\def\HBM{\mathcal{H}}

\begin{document}

\title[Magic distances in twisted bilayer graphene]{Magic distances in twisted bilayer graphene}


\author[1,2]{\fnm{Antonio} \sur{Palamara}}\email{antonio.palamara@unical.it}

\author*[1,2]{\fnm{Michele} \sur{Pisarra}}\email{michele.pisarra@unical.it}

\author[1,2]{\fnm{Antonello} \sur{Sindona}}\email{antonello.sindona@fis.unical.it}

\affil[1]{\orgdiv{Dipartimento di Fisica}, \orgname{Università della Calabria}, \orgaddress{\street{Via P. Bucci, Cubo 30C}, \city{Rende (CS)}, \postcode{87036}, \country{Italy}}}

\affil[2]{\orgdiv{INFN, sezione LNF, Gruppo Collegato di Cosenza}, \orgaddress{\street{Via P. Bucci, Cubo 30C}, \city{Rende (CS)}, \postcode{87036}, \country{Italy}}}


\abstract{Twisted bilayer graphene exhibits isolated, relatively flat electronic bands near charge neutrality when the interlayer rotation is tuned to specific \textit{magic} angles.
These small misalignments, typically below 1.1°, result in long-period moir\'{e} patterns with anomalous electronic properties, posing severe challenges for accurate atomistic simulations due to the large supercell sizes required.
Here, we introduce a framework to map arbitrarily stacked graphene bilayers, characterized by specific rotation angles corresponding to precise interplanar distances, onto an equivalence class represented by \textit{magic-angle} twisted bilayer graphene.
Using a continuum model, we derive the equivalence relation defining this class and extend its implementation to tight-binding approaches.
We further explore the applicability of this mapping within density functional theory, demonstrating that the \textit{magic-angle physics} can be efficiently studied using twisted bilayer graphene configurations with larger stacking angles and computationally manageable supercell sizes.
This approach offers a pathway for \textit{ab~initio} investigations into unconventional topological phases and emergent excitations in the low-energy quasi-flat bands of twisted bilayer materials.}




\maketitle

\section{Introduction}\label{sec1}

Twisted bilayer graphene~(TBG) is a unique structure formed by stacking two graphene layers with a controlled interlayer distance and rotational mismatch, resulting in a compelling interplay between electronic and geometric effects.
At specific \emph{magic} twist angles, the TBG lattice  produces large moir\'{e} patterns, giving rise to strongly correlated phenomena such as unconventional Mott insulating, Chern insulating, and superconducting phases~\cite{cao2018unconventional, cao2018correlated, yankowitz2019tuning, lu2019superconductors, sharpe2019emergent, Park2021}.
These exotic behaviors arise from nearly flat electronic bands with vanishing group velocity at the Fermi level, as predicted by numerous theoretical studies~\cite{Bistritzer, PhysRevB.81.245412, watson2023bistritzer, PhysRevB.86.125413, PhysRevB.86.155449, PhysRevLett.99.256802, 10.21468/SciPostPhys.7.4.048, PhysRevLett.122.106405, morell2010flat, PhysRevLett.125.214301} and confirmed by experimental observations~\cite{xie2019spectroscopic, kerelsky2019maximized, jiang2019charge, li2010observation, KimPNAS2017, PhysRevLett.109.196802, tong2022spectroscopic, choi2019electronic, lisi2021observation}.

The discovery of \emph{magic-angle}~(MA) TBG has further broadened the already rich field of van der Waals~(VdW) heterostructures~\cite{Geim_Grigorieva}, introducing the interlayer twist angle $\theta$ as an additional degree of freedom in their assembly.
This breakthrough facilitates the on-demand tuning of the electronic, mechanical, and thermal properties of two-dimensional~(2D) materials via VdW interactions, while significantly advancing the rapidly evolving field of twistronics~\cite{Twistronics1, MATBG}.
However, $\theta$ is not the only geometric parameter that regulates the formation of quasi-flat bands.
The interlayer distance $d$ also plays a pivotal role, as vertical pressures in the gigapascal range can induce weakly dispersive electronic states in non-MATBG systems.~\cite{Chittari_2019, PhysRevB.99.045423, PhysRevB.98.085144, D1NR00220A, PhysRevB.101.155405}.

In the simplest picture~\cite{Bistritzer, PhysRevB.81.245412, watson2023bistritzer}, the emergence of quasi-flat bands stems from the interplay of two energy scales.
One is determined by $\theta$ and involves the electronic states over the moir\'{e} superlattice sites.
The other arises from the coupling between the lattice sites of the TBG layers and is dependent on $d$.
The characteristic features of MATBG, however, are typically observed at small twist angles~($\theta{\sim}1.1^\circ$), requiring supercells of large sizes~($\sim 10^4$ atoms).
This requirement presents substantial challenges for accurate modeling due to the significant computational resources required~\cite{PhysRevB.99.195419}. 

On these premises, we here investigate the existence of an infinite set of crystal~(and potentially quasi-crystal) structures, defined by various values of $\theta$ corresponding to precise values of $d$, all of which belong to an equivalence class with MATBG as a representative member.
The mapping between the elements of this class enables a detailed exploration of the MA quasi-flat-band physics, focusing on identifying the most suitable geometry within the class for atomistic simulations.
While other mappings have been developed to effectively reduce the unit cell size of a TBG moir\'{e} crystal~\cite{PhysRevB.104.075144, 10.21468/SciPostPhys.11.4.083, PhysRevLett.119.107201, prbnanoribbon, PhysRevX.8.031087, PhysRevResearch.1.013001, PhysRevB.96.075311}, they are typically derived by operating on the free parameters of specific model approaches. 
In contrast, our formulation is based on two experimentally observable quantities, such as interlayer distance and/or twist angle, making it independent of the underlying theory.

\section{Results}\label{sec2}
\subsection{The equivalence class of MATBG\label{EqCMsec}}
The continuum model~(CM) provides a foundational approach for analyzing and predicting the emergence of quasi-flat bands in TBG with a stacking angle $\theta > 0$. 
As detailed in section I of the SI,
this behavior originates from the Bistritzer-MacDonald~(BM) Hamiltonian~\cite{Bistritzer, PhysRevB.81.245412, watson2023bistritzer}: 
\begin{equation}
\HBM = {\hbar}{\kth}{\vF}  \tilde{\mathcal{H}}_{0} + {\omega} \tilde{\mathcal{H}}_{\msc{t}} ={\hbar}{\kth}{\vF} \left( \tilde{\mathcal{H}}_{0} + {\gamma} \tilde{\mathcal{H}}_{\msc{t}} \right),
\label{MC}
\end{equation}
which introduces two characteristic energies. 
The first one, ${\hbar}{\kth}{\vF}$, is the  single-layer-graphene quasiparticle energy at the characteristic momentum $k_{\theta}$,
 which establishes the energy scale of the dimensionless \emph{unperturbed} Hamiltonian $\tilde{\mathcal{H}}_{0}$. 
This term describes the \emph{uncoupled} graphene layers, whose equivalent Dirac cones exhibit a Fermi velocity $\vF$ and are separated by the wave vector $\kth = 2K_{\msc{d}} \sin(\theta/2)$, where $K_{\msc{d}}$ represents the distance from the center to a corner of the hexagonal first Brillouin zone~({\fbz}) of graphene. 
The second is the interlayer hopping energy, ${\omega}{=}{\hbar}{\kth}{\vF}{\gamma}$~\cite{Bistritzer, PhysRevB.81.245412, watson2023bistritzer, PhysRevB.86.125413}, which is assumed to depend only on the interlayer spacing $d$ and not on the specific atomic stacking within the moir\'{e} unit cell~\cite{PhysRevB.103.205413, PhysRevB.90.155451, Wijk_2015, dai2016twisted, PhysRevB.96.075311}. 
This energy scale determines the strength of the dimensionless \emph{interlayer} potential $\tilde{\mathcal{H}}_{\msc{t}}$. 
The distinctive feature of the BM Hamiltonian is that the underlying physics of the system is governed by a single dimensionless parameter, $\gamma$, apart from an overall energy scaling factor. 
This formulation implies that varying $\omega$ and $\theta$ while keeping $\gamma$ fixed does not alter the eigenstates of $\HBM$.
The corresponding eigenvalues are either expanded or compressed by a factor of $\hbar\kth\vF$.
Numerical calculations reveal the emergence of quasi-flat energy bands with vanishing Fermi velocity for appropriate choices of $\gamma$ and $\omega$.
Specifically, $\gamma{=}\gaM{=}3^{-1/2}$ and $\omega{=}\wM{=}0.11$~eV correspond to a sequence of MAs with a maximum value of $\thM{\sim}1.05^{\circ}$~\cite{Bistritzer, PhysRevB.81.245412, watson2023bistritzer}.

Let us now denote by $\mathcal{O}_{\msc{bm}}$ the set of all Hamiltonians of the BM type, given by Eq.~(\ref{MC}) and depending on $\omega$ and $\theta$.
Consider two generic elements $\mathcal{H}_{1}{=}\HBM(\omega_{1},\theta_{1})$ and $\mathcal{H}_{2}{=}\HBM(\omega_{2},\theta_{2})$ in $\mathcal{O}_{\msc{bm}}$.
We may introduce the equivalence relation:
\begin{equation}
\mathcal{H}_{1}\underset{\gamma}{\sim }\mathcal{H}_{2} \iff
\frac{\omega _{1}}{\hbar k_{\theta _{1}}\vF}
=\frac{\omega_{2}}{\hbar k_{\theta _{2}}\vF}
=\mathcal{\gamma }
\label{eqclass}
\end{equation}
which partitions $\mathcal{O}_{\msc{bm}}$ into equivalence classes of the form:
\begin{equation}
[\mathcal{H}]_{\gamma} \equiv \{
 \mathcal{H}\in \mathcal{O}_{\msc{bm}}:
 \mathcal{H}_{\gamma}\underset{\gamma}{\sim}\mathcal{H}
 \}, \label{eqrel}
\end{equation}
where $\mathcal{H}_{\gamma}$ serves as a representative of the $\gamma$ class.
If $\mathcal{H}_{1}$ and $\mathcal{H}_{2}$ belong to $[\mathcal{H}]_{\gamma }$, they share the same eigenstates and exhibit \emph{homothetic} spectra, which differ only by a scaling factor.
In particular, the spectrum of $\mathcal{H}_{1}$ is obtained from the spectrum of $\mathcal{H}_{2}$ by multiplication with:
\begin{equation}
s^{1}_{2}=\frac{k_{\theta_{1}}}{k_{\theta_{2}}}=\frac{\sin(\theta_{1}/2)}{\sin(\theta_{2}/2)}.
\label{eqscale}
\end{equation}
Among the various $\gamma$ classes in $\mathcal{O}_{\msc{bm}}$, we can identify a distinct equivalence class, denoted by ${\gamma}{=}\gaM$, whose representative element corresponds to one of the previously described MA Hamiltonians, such as the one characterized by $\omega{=}\wM$ and $\theta{=}\thM$.
Consequently, all Hamiltonians within $[\mathcal{H}]_{{\gaM}}$ exhibit the same number of energy bands with vanishing group velocity at the Fermi level.
We recall that the interlayer coupling decreases as the interlayer distance $d$ increases. This observation motivates the assumption of a bijective function $F(d)$ that relates the hopping energy to the interlayer distance:
\begin{equation}
\omega(d)=\omega_0 F(d),
\label{omega(d)}
\end{equation}
where $F(d_{0}){=}1$ at a reference value $d_{0}$, ensuring $\omega_0{=}\omega(d_0)$.
Let us consider the equivalence class $\gamma$, represented by a twist angle $\theta_{0}$ at the interlayer distance $d_{0}$.
By the defining relation~(\ref{eqclass}) and the bijectivity of $F$, there exists a unique interlayer distance:
\begin{equation}
d_{\gamma}(\theta)= F^{-1}\left(\frac{\omega}{\omega_0}\right)= F^{-1}\left(\frac{\kth}{k_{\theta_{0}}}\right),
\label{dt}
\end{equation}
for any other twist angle $\theta$, which corresponds to a BM Hamiltonian within the same equivalence class.
Let us now return our focus to the equivalence class of MATBG, represented by the Hamiltonian $\mathcal{H}_{{\gaM}}$ with geometric parameters $\thM$ and $\wM$.
We may set $\theta_{0}{=}\thM$ and associate $\wM$ with the nominal interlayer distance $d_{0}{=}3.349$~{\AA}, characteristic of bilayer graphene or graphite.
For any twist angle $\theta$, we can then determine the corresponding \emph{magic} interlayer distance $d_{\gaM}(\theta)$.
These two parameters together define a BM Hamiltonian within $[\mathcal{H}]_{\gaM}$, which hosts the same number of bands with vanishing Fermi velocity as the MATBG Hamiltonian.

The construction presented thus far is valid at the CM level, whose limits are outlined in 
Supplementary Section II.
Relaxation effects can be straightforwardly incorporated into the framework by redefining the hopping energy $\omega$ in Eq.~(\ref{MC}), as also discussed in Supplementary Section.
Nonetheless, $d$ and $\theta$ remain the two experimentally tunable geometric parameters that uniquely determine the configuration of a TBG structure, up to symmetry operations.
For this reason, it should always be feasible to construct an equivalence class for the electronic states near the Fermi level, regardless of the defining model.
This conjecture leads to two notable consequences.
(i) In experiments involving TBG samples at small twist angles, a group of quasi-flat bands at the Fermi level should emerge through the application and tuning of external pressure, as confirmed by previous studies~\cite{Chittari_2019, PhysRevB.99.045423, PhysRevB.98.085144, D1NR00220A, PhysRevB.101.155405}.
These compressed bands would realize a mapping of the MA quasi-flat bands, enabling an experimental validation of the equivalence relation defined above.
(ii) A comprehensive characterization of the flat-band physics of MATBG can be achieved using an equivalent TBG system with a twist angle larger than the MA. Such a system contains fewer atoms per unit cell, making it computationally less demanding while preserving the same features as MATBG on an expanded energy scale.

\subsection{Magic distances and equivalent band structures in the tight binding formalism\label{TBsec}}
\begin{figure}[htb]
\centering
\includegraphics[width=0.99\linewidth]{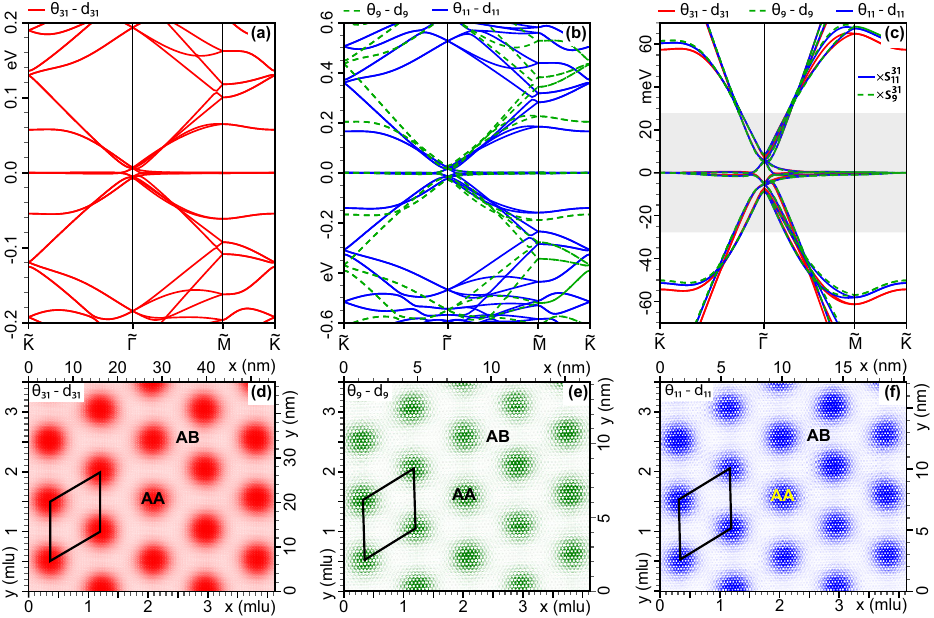}
\caption{(a), (b) TB band structures of TBG lattices with geometric parameters: (a) $\theta_{31}{=}1.050^{\circ}$-$d_{31}{=}3.349$~{\AA} (representing the MA condition), (b) $\theta_{9}{=}3.481^{\circ}$-$d_{9}{=}2.7329$~{\AA}~(green dashed lines), and  $\theta_{11}{=}2.876^{\circ}$-$d_{11}{=}2.8304$~{\AA}~(blue lines).
(c)~Compressed band structures from~(b), scaled by factors of $s^{31}_{9}{=}0.3017$~(green dashed lines) and $s^{31}_{11}{=}0.3652$~(blue lines), superimposed onto the MATBG bands~(red lines) from~(a); the three plots coincide in the $\pm20$meV energy range, highlighted by the shaded region.
All band dispersions are plotted along the $\GKMGt$ path with the Fermi energy set to zero.
(d)-(f) Projected local density near the Fermi level for: (d)~MATBG~(red spots), (e)~TBG with $\theta_{9}$-$d_{9}$~(green spots), and (f)~TBG with $\theta_{11}$-$d_{11}$~(blue spots).
The in-plane coordinates are given in both nm and moir\'{e} lattice units~(mlu).
The black parallelogram represents the moir\'{e} unit cell of the respective TBG system.
\label{fig1}}
\end{figure}

Our next objective is to numerically assess the existence of an equivalence class for MATBG and establish a suitable functional form for Eqs.~(\ref{omega(d)} and \ref{dt}).
As a validation tool, we employ a tight binding~(TB) approach grounded in a quadratic Hamiltonian, featuring diagonal on-site energies and off-diagonal inter-site hopping potential matrix elements.
This framework enables a systematic examination of the electronic properties of TBG structures across varying values of $\theta$ and $d$.
The TB calculations were carried out using the TBPLas package~\cite{LI2023108632}, which employs the Slater-Koster~(SK) formalism~\cite{PhysRev.94.1498,PhysRevB.98.081410, PhysRevB.98.235137}, as detailed in the Methods section.
We specifically selected TBG geometries where the mismatch angle between the two graphene layers results in commensurate arrangements~\cite{PhysRevB.81.161405, PhysRevB.81.165105}.
In this scenario and using AA-stacked bilayer graphene as a starting point, the twist angle is given by:
\begin{equation}
\cos(\theta_m)=\frac{3m^2+3m+1/2}{3m^2+3m+1},
\label{eq2}
\end{equation}
where $m$ is an integer.
Applying the SK parameterisation at the equilibrium interlayer distance $d_{0}{=}3.349$~{\AA}~\cite{NatCom.Shi2020}, the TB Hamiltonian produces four nearly flat bands with vanishing Fermi velocity at $\theta_{31}{=}1.050^{\circ}$. 
This is illustrated in Fig.~\ref{fig1}a, which shows the band structure of MATBG~(red lines) along the $\GKMGt$ borders of the irreducible {\fbz} of the corresponding moir\'{e} lattice~(see also Supplementary Figure 1 in Supplementary Section I).

The subsequent step is to construct the MATBG equivalence class for the TB method.
To this end, we adjusted the $d$ parameter to various interlayer distances while keeping the twist angle $\theta_{m}$ fixed.
During the process, we searched for the band structure with nearly non-dispersive states at the Fermi level that most closely resemble the MA bands when scaled by the factor $s^{31}_{m}$, cf.~Eq.(\ref{eqscale}).
For each tested $\theta_{m}$, we identified a well-converged \emph{magic distance} $d_{m}$ that satisfies this criterion. 
To explicitly illustrate the procedure, we discuss the results for two twist angles greater than $\theta_{31}$, namely, $\theta_{9}{=}3.481^{\circ}$ and $\theta_{11}{=}2.876^{\circ}$.
The search returned the magic distances $d_{9}{=}2.7329$~{\AA} for $\theta_{9}$ and $d_{11}{=}2.8304$~{\AA} for $\theta_{11}$.
Fig.~\ref{fig1}b presents the band structures corresponding to $\theta_{9}$-$d_{9}$~(dashed green lines) and $\theta_{11}$-$d_{11}$~(blue lines), which appear to differ merely by an overall scaling factor.
Furthermore, the MATBG spectrum in Fig.~\ref{fig1}a~(red lines) bears a striking resemblance to the TBG spectra in Fig.~\ref{fig1}b, although on a different energy scale.
We then applied the scaling factors $s^{31}_{9}{=}k_{\theta_{31}}/k_{\theta_{9}}{=}0.3017$ and $s^{31}_{11}{=}k_{\theta_{31}}/k_{\theta_{11}}{=}0.3652$ to the band systems from Fig.~\ref{fig1}b and compared them with the MATBG bands from Fig.~\ref{fig1}a.
The scaled dispersions from the two non-MATBG structures practically coincide with the MATBG dispersions in their low-energy regions, within ${\sim}0.04$~eV around the Fermi level, as shown in Fig.~\ref{fig1}c.

Strictly speaking, the equivalence class $\gaM$ is well-defined within this energy range, which establishes the limits of validity for the CM.
Nonetheless, we propose extending the definition~(\ref{eqclass}) to the TB level, asserting that two TB Hamiltonians are equivalent if their energy spectra differ only by an overall scaling factor in the CM limit.
In this context, the low-energy bands corresponding to $\theta_{9}$-$d_{9}$ and $\theta_{11}$-$d_{11}$ are effectively mapped onto the MATBG quasi-flat bands.
To fully validate this relation, it is also necessary to confirm that the spatial distributions of the band states for the three equivalent TBG configurations exhibit closely similar profiles within the CM range.
To this end, we employed the TB propagation method~\cite{LI2023108632} to investigate the local electronic densities of the TBG systems, projected onto the quasi-flat-band range characteristic of MATBG.
Figs.~\ref{fig1}d-f reveal that these densities are strongly localized around the AA stacking regions and rapidly vanish near the AB stacking regions.
This result aligns with previous studies~\cite{KimPNAS2017, li2010observation, PhysRevLett.109.196802, tong2022spectroscopic} and, more importantly, confirms that the eigenstates corresponding to the quasi-flat bands within the same equivalence class exhibit highly similar profiles and identical localization properties.

\begin{figure}[h]
\centering
\includegraphics[width=0.7\textwidth]{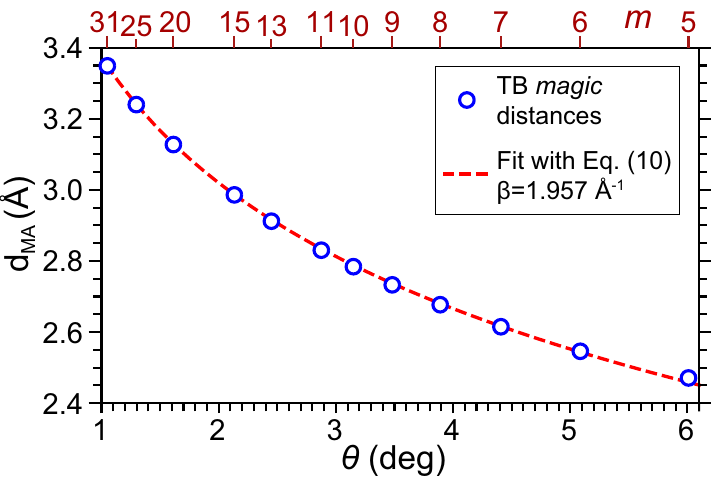}
\caption{Magic distance $d_{\msc{ma}}$ for the twist angle $\theta$.
The blue circled points label the $\theta_{m}$-$d_{m}$ data obtained via the TB method, with the top horizontal axis indicating $m{=}5$-$31$, cf. Eq.~(\ref{eq2}).
The red dashed line corresponds to the function in Eq.~(\ref{dma}), with the best-fit parameter $\beta{=}1.957$~{\AA}$^{-1}$.}
\label{figdma}
\end{figure}

To complete the picture, we applied the equivalence mapping procedure to a series of TBG systems with twist angles ranging from $\theta_{31}{=}1.050^{\circ}$ to $\theta_{5}{=}6.009^{\circ}$. 
Fig.~\ref{figdma} displays the interlayer magic distances $d_{m}$, at which nearly non-dispersive bands with identical localization properties were obtained, as a function of $\theta_m$, for $m{=}5$-$31$~(blue circled points). 
The plot reveals an inverse correlation between $\theta_m$ and $d_m$: as the twist angle increases, the corresponding magic distance decreases. This behavior suggests that realizing  magic conditions at a twist angle greater than the original magic angle requires a reduced interlayer spacing, which enhances the interlayer coupling. 
Such an effect can be experimentally achieved through applied external pressure.
In the TB framework, a reduction in interlayer spacing corresponds to increasing the interlayer hopping amplitudes, consistent with enhanced overlap between the atomic wavefunctions of the  two graphene layers. 
Given the characteristic exponential decay of the electronic density in the out-of-plane direction for 2D materials, it is reasonable to assume a similar dependence of the interlayer hopping terms. Accordingly, we adopt an exponential scaling law for the function $F(d)$ in Eq.~(\ref{omega(d)}), which means setting:
\begin{equation}
\omega(d)=\omega_{0}F(d)=\omega_{0}e^{-\beta(d-d_{0})}, \quad {d}\leq{d_{0}}.
\label{wbeta}
\end{equation}
Here $d_0{=}d_{31}{=}3.349$~{\AA}, $\theta_0{=}\theta_{31}{=}1.05^{\circ}$, and $\omega_0{=}\omega_{\msc{MA}}{=}0.11$~eV serve as reference values, while the inverse damping length $\beta$ governs the attenuation of $\omega(d)$.
Operating within the equivalence class of MATBG we impose the condition:
\begin{equation}
\frac{\omega(d_{\msc{ma}})}{\hbar k_{\theta}\vF}
{=}\frac{\omega_{\msc{ma}}}{\hbar k_{\theta _{0}}\vF},
\end{equation}
to establish a direct relation between the TB magic distances $d_{\msc{ma}}$ and the commensurate twist angles $\theta$.
This analysis yields the function:
\begin{equation}
d_{\msc{ma}}(\theta)=d_{0}-\frac{1}{\beta}\ln\left[\frac{\sin({\theta}/{2})}{\sin({\theta_{0}}/{2})}\right], \label{dma}
\end{equation}
which we fitted to the $\theta_{m}$-$d_{m}$ data, to determine the optimal value of $\beta{=}1.957~\text{\AA}^{-1}$.
The resulting fit, represented by the red dashed line in Fig.~\ref{figdma}, offers an excellent approximation to the bijective function proposed in Eq.~(\ref{dt}).

The procedure outlined above can be extended to generate all equivalence classes defined in Eq.~\ref{eqclass}. In Supplementary Section III this is exemplified through the class whose representative element is a TBG structure with $\theta_{15}=2.135^{\circ}$ and the nominal interlayer distance $d_0=3.345$~\AA.
We have thus presented strong evidence that the MA condition corresponds to a well-defined equivalence class at the TB level. This supports our initial observation that the equivalence class of MATBG can be fully constructed by modifying the model-independent parameters $\theta$ and $d$, as further detailed in Supplementary Section IV. 

We conclude this section observing that many TB calculations of TGB structures explicitly incorporate structural relaxation effects~\cite{PhysRevB.106.115410,PhysRevResearch.2.043127,PhysRevB.99.195419}. In principle, the equivalence class approach outlined above within a TB framework could be extended to account for out-of-plane corrugation. In this generalized scenario a given twist angle would be associated with both a magic interlayer distance and at least one corresponding \emph{magic corrugation amplitude}. Such an extension would also require a careful re-definition of the out-of-plane hopping term $V_{pp\sigma}$ in Eq.~\ref{Vpps}. A detailed discussion of the role of corrugation within the equivalence class framework, as well as the impact of a model corrugation~\cite{PhysRevB.90.064101} on the quasi-flat bands at the TB level are provided in Supplementary Section V. In contrast, in-plane reconstruction effects, which are also expected when full relaxation of TBG structure are carried out \cite{PhysRevB.106.115410,PhysRevResearch.2.043127,PhysRevB.99.195419}, cannot be readily incorporated, as they simultaneously affect both the in-plane and the out-of-plane TB hopping amplitudes.  
Nonetheless, the core message of our study remains unaffected: geometric tuning provides a practical and general route to achieving magic-angle-like conditions at larger twist angles, thereby enabling ab initio methods, as demonstrated below using density functional theory~(DFT).

\subsection{Equivalence-class approach and DFT\label{DFTsec}}
The BM model finds formal justification at the DFT level~\cite{PhysRevB.107.155403}. 
It is therefore reasonable to utilize the equivalence-class concept within this \textit{ab~initio} framework to establish a connection between the near-Fermi-energy features of TBG systems and the MA condition.
To this end, we carried out state-of-the art DFT calculations of TBG structures using the Perdew-Burke-Ernzerhof~(PBE) exchange-correlation functional~\cite{PBE_xc} and the Projector Augmented Wave method~\cite{PhysRevB.50.17953}, as implemented in the Quantum Espresso package ~\cite{Giannozzi_2009,Giannozzi_2017}.  

To ensure computational feasibility, we focussed on commensurate lattices with twist angles $\theta_{5}{=}6.009^{\circ}$, $\theta_{6}{=}5.086^{\circ}$, $\theta_{7}{=}4.408^{\circ}$, $\theta_{8}{=}3.890^{\circ}$, and $\theta_{9}{=}3.481^{\circ}$.
These angles were selected as they are considerably larger than the MA while remaining within the range of validity of our scaling procedure established at the TB level.
The corresponding configurations required ground-state calculations involving $364$, $508$, $676$, $838$, and $1084$ atoms per unit cell, respectively.
We opted against directly evaluating the electronic properties of MATBG due to the very large size of its unit cell. Indeed, the twist angle $\theta_{31}$, which satisfies the MA condition at the equilibrium interlayer distance $d_{0}$ within the TB approach, results in a commensurate TBG lattice containing $11908$ atoms per unit cell.
Such a system renders DFT-based band-structure calculations highly demanding, significantly constraining the achievable accuracy even with advanced high-performance computing resources~\cite{PhysRevB.99.195419}.
Moreover, it is not guaranteed that the MA condition holds precisely at $\theta_{31}$ within the DFT-PBE framework.

\begin{figure}[t]
\centering
\includegraphics[width=0.80\textwidth]{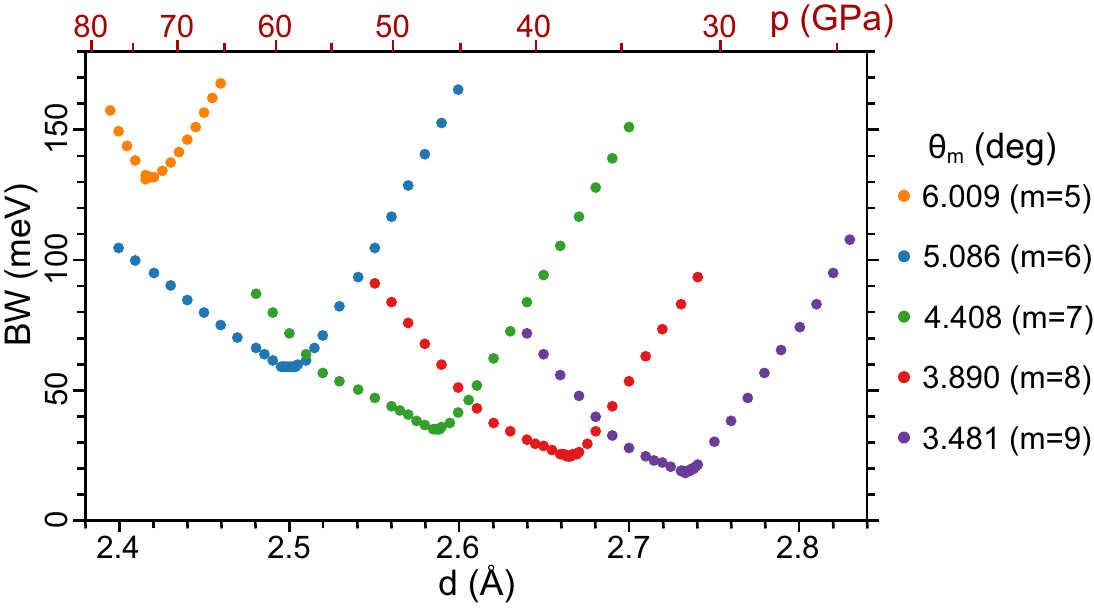}
\caption{Total BW of the four bands closest to the Fermi level \textit{vs.} interlayer separation $d$, obtained from our DFT calculations for commensurate TBG systems with twist angles from $\theta_{5}$ to $\theta_{9}$, cf. Eq.~(\ref{eq2}).
The top horizontal axis indicates the applied pressure required to achieve each value of $d$, as obtained from the DFT calculated forces acting on the atoms~\cite{Ares.AdvFunMat2019}.
\label{figBW}}
\end{figure}

With the objective of identifying four non-dispersive bands at the Fermi level under the MA mapping, we conducted a series of self-consistent ground-state calculations, where the interlayer distance $d$ was systematically varied, and the total bandwidth~(BW) of the Kohn-Sham~(KS) states nearest to the Fermi level was monitored throughout. 
It is worth mentioning that this particular way of searching for magic-like conditions is not unique. In principle, one could instead minimize the group velocity, $\mathrm{v}_{\msc{g}}$, or maximize the effective masses, $m^*$, of the bands close to the Fermi level. In practice, however, the evaluation of $\mathrm{v}_{\msc{g}}$ or $m^*$ requires a very dense Brillouoin Zone sampling, which dramatically increases the computational cost, and can be, in general, very challenging in systems like TBG, characterized by huge moiré unit cells~\cite{ARNOLD202496}.    
Figure~\ref{figBW} illustrates  the different BWs corresponding to the selected twist angles as functions of $d$.
Each BW curve exhibits a distinct minimum, which we designate as the \emph{DFT magic distance}.
Beyond this point, the BW increases sharply with larger $d$ values and rises more gradually at smaller $d$ values.

Notably, for the twist angles  $\theta_6$ to $\theta_9$ the BW takes on very small values, on the order of a few tens of meV, across a relatively broad range of interlayer distances, approximately of $0.2$–$0.4$~{\AA}. Within this \emph{minimum region} any value of $d$ presents magic-angle-like conditions  for the four bands closest to the Fermi level.
In contrast, while the BW at $\theta_5$ still exhibits a clear minimum, its overall shape differs noticeably from other BW curves.
Figure~\ref{figBW} also reveals two key trends as the selected twist angle decreases.
(i) The minimum BW shifts to occur at increasingly larger interlayer distances, corresponding to decreasingly lower applied pressures, as indicated on the top horizontal axis.
(ii) The minimum BW progressively decreases, with values at twist angles $\theta_8$ and $\theta_9$ dropping well below $30$~meV, which is roughly the expected BW predicted by earlier DFT calculations on MATBG~\cite{PhysRevB.99.195419}.
These results are consistent with the expectation that, as the twist angle approaches the MA under equilibrium interlayer distance and zero applied pressure, the BW should ideally reach its minimum possible value.

\begin{figure}[thb]
\centering
\includegraphics[width=0.99\textwidth]{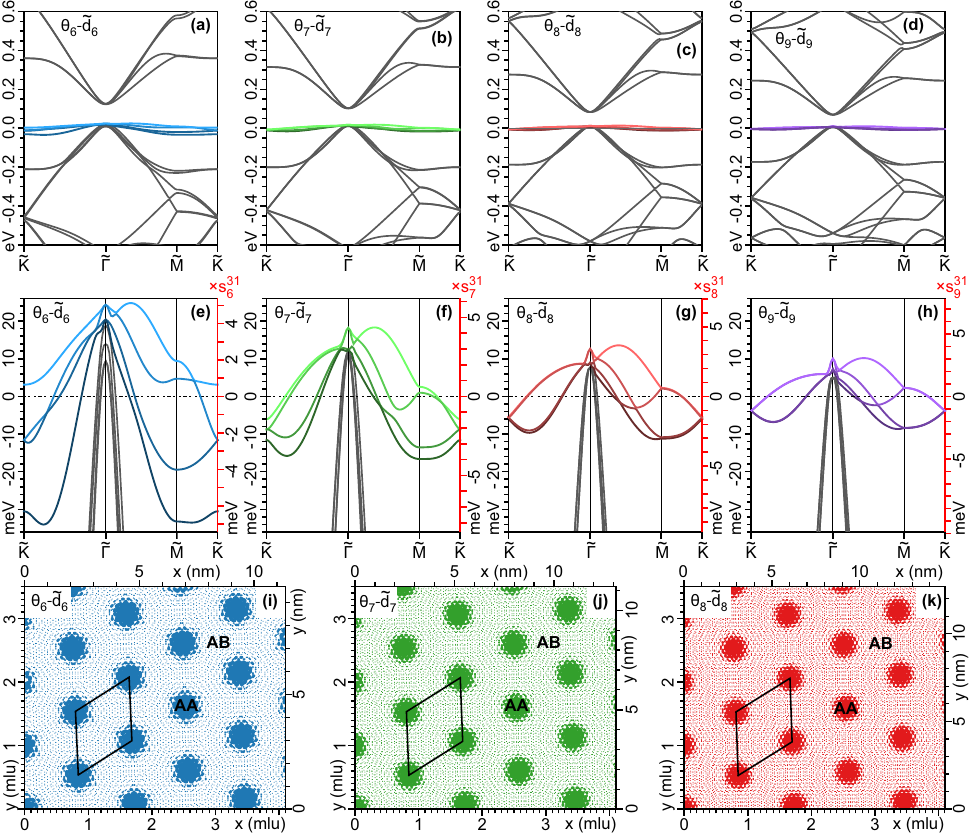}
\caption{(a)-(h) DFT band energies~(left vertical axis) and (e)-(h) compressed band energies~(right vertical axis) along the $\GKMGt$ path for TBG lattices with geometric parameters: (a), (e) $\theta_{6}{=}5.086^\circ$-$\tilde{d}_{6}{=}2.497$~{\AA}, (b), (f) $\theta_{7}{=}4.408^\circ$-$\tilde{d}_{7}{=}2.587$~{\AA}, (c), (g) $\theta_{8}{=}3.890^\circ$-$\tilde{d}_{8}{=}2.665$~{\AA}, and (d), (h) $\theta_{9}{=}3.481^\circ$-$\tilde{d}_{9}{=}2.733$~{\AA}, corresponding to the scaling factors: (e) $s^{31}_{6}{=}0.2065$, (f) $s^{31}_{7}{=}0.2383$, (g) $s^{31}_{8}{=}0.2700$, and (h) $s^{31}_{9}{=}0.3017$.
The Fermi energy is set to zero.
The four quasi-flat bands used to evaluate the BW in Fig~\ref{figBW} are highlighted in colour, while other dispersive bands are shown in grey.
(i)-(k) Projected local density of states corresponding to the quasi-flat bands in (a)-(c), respectively.
The in-plane coordinates are given in both nm and mlu, with the black parallelogram denoting the moir\'{e} unit cell of the respective TBG system, as in Fig~\ref{fig1}d-f.
\label{figbdft}}
\end{figure}

A refined search within the region of minimum BW identified the magic distances $\tilde{d}_{6}{=}2.497$~{\AA}, $\tilde{d}_{7}{=}2.587$~{\AA}, $\tilde{d}_{8}{=}2.665$~{\AA}, and $\tilde{d}_{9}{=}2.733$~{\AA}, associated to the twist angles $\theta_{6}$ through $\theta_{9}$, respectively.
At these interlayer separations, the four bands closest to the Fermi level exhibit minimal dispersion, categorising them as quasi-flat bands.
Figures~\ref{figbdft}a-h display these quasi-flat bands, coexisting with other dispersive bands within the same energy range, 
along the $\GKMGt$ path,
highlighting the following recurring features for TBG lattices with geometric parameters from $\theta_7$-$\tilde{d}_{7}$ to $\theta_{9}$-$\tilde{d}_{9}$.

(i) There is a marked difference in the dispersion of the upper and lower quasi-flat bands. The \emph{two highest} quasi-flat bands display similar dispersions near the $\tilde{\Gamma}$ point, forming a peak above the Fermi level. They overlap along $\tilde{K}\tilde{\Gamma}$ and $\tilde{M}\tilde{K}$ but remain distinct along $\tilde{\Gamma}\tilde{M}$.
The \emph{two lowest} quasi-flat bands also show broad maxima at $\tilde{\Gamma}$ and approach degeneracy along $\tilde{K}\tilde{\Gamma}$ and $\tilde{M}\tilde{K}$, with minor differences persisting along $\tilde{\Gamma}\tilde{M}$.
Such \emph{asymmetry} is not found at the TB level.

(ii) A set of parabolic dispersive bands crosses the Fermi level and converges at a single maximum at $\tilde{\Gamma}$, where they touch the quasi-flat bands.
Another set of parabolic-like states lies above the quasi-flat bands, with a common minimum at $\tilde{\Gamma}$ that ranges from approximately ${\sim}120$~meV for $\theta_6$-$\tilde{d}_{6}$ to ${\sim}70$~meV for $\theta_9$-$\tilde{d}_{9}$.

(iii) The dispersive bands in the wide-range plots in Figs.\ref{figbdft}a-d show strong similarities with one another (they roughly differ by a scaling factor). Remarkably they closely resemble the dispersive bands obtained in the TB calculations (cf. Fig.\ref{fig1}a-c) with the important distinction of a gap opening between the quasi-flat bands and the dispersive bands above the Fermi level.

(iv)The band structure for the geometric parameters $\theta_{6}$-$\tilde{d}_{6}$, shown in Fig~\ref{figbdft}e, exhibits notable differences
compared to those in Figs.~\ref{figbdft}b-d. Moreover, although not explicitly presented here, the band dispersion for TBG with a twist angle $\theta_{5}$ and an interlayer distance of $\tilde{d}_{5}{=}2.4170${\AA}, corresponding to the minimum of the associated BW curve in Fig~\ref{figBW}, displays markedly distinct characteristics, seemingly uncorrelated with the other cases. 
 This observation aligns with the detection of a wide BW for $\theta_{5}$ at the TB level, indicating that $\theta{\sim}6^{\circ}$ exceeds the upper limit of applicability for our mapping procedure. 

As another analogy to the TB-based derivation of the MATBG equivalence class, Figs.~\ref{figbdft}e-g present the local electron density projected onto the four quasi-flat bands for the cases $\theta_{6}$-$\tilde{d}_{6}$, $\theta_{7}$-$\tilde{d}_{7}$, and $\theta_{8}$-$\tilde{d}_{8}$. 
Consistent with the analysis in Fig~\ref{fig1}, these spatial distributions demonstrate that the quasi-flat-band states are strongly localized within the AA stacking regions, characterized by pronounced density peaks, whereas the AB stacking regions exhibit density minima.

Thus, the TBG lattices analyzed here using DFT exhibit notable similarities to MATBG, as discussed at the TB level.
All these systems are characterized by four nearly flat bands around the Fermi level, with electrons predominantly localized within the AA stacking regions of the moir\'{e} lattice.
While precise scaling factors for transforming one set of quasi-flat bands  into another remain unavailable at the DFT level, applying the scaling factors $s^{31}_{6}{=}0.2065$, $s^{31}_{7}{=}0.2383$, $s^{31}_{8}{=}0.2700$, and $s^{31}_{9}{=}0.3017$ from Eq.~(\ref{eqscale}) to the DFT band structures results in total widths of the four bands comparable to, or slightly smaller than, those of MATBG at the TB level, being approximately $12$~meV, as shown in Fig~\ref{fig1}c. 
This outcome is apparent from the right vertical axis in Figs.~\ref{figbdft}a-d, where differences in band dispersion diminish to below ${\sim}1$~meV for the $\theta_{8}$-$\tilde{d}_{8}$ and $\theta_{9}$-$\tilde{d}_{9}$ cases.

Comparable features are observed in TBG systems with interlayer distances $d$ close to the magic values $\tilde{d}_{m}$~($m{=}6,{\ldots},9$), identified within the minimum regions of Fig~\ref{figBW}, although the  detailed dispersion of the quasi-flat band exhibit some variations. A specific case for $m=7$ is discussed in Supplementary Section VI.
It is also important to note that for stacking angles significantly larger than the MA, the dispersion of the quasi-flat bands depends on the specific construction of the moir\'{e} unit cell, particularly the placement of the rotation axis, as shown in Supplementary Section VII.

We interpret the small energy differences in the quasi-flat bands as an inherent measure of error when reproducing the DFT features of MATBG through the equivalence-class approach.
From this analysis, we infer that the TBG systems considered here belong to \emph{nearby} equivalence classes to that of MATBG, characterised at the CM level by an interaction parameter  $\gamma{=}\gamma_{\msc{ma}}+\delta \gamma$, with ${\delta\gamma/\gaM}{\ll}1$.

\begin{figure}[tbh]
\centering
\includegraphics[width=0.7\textwidth]{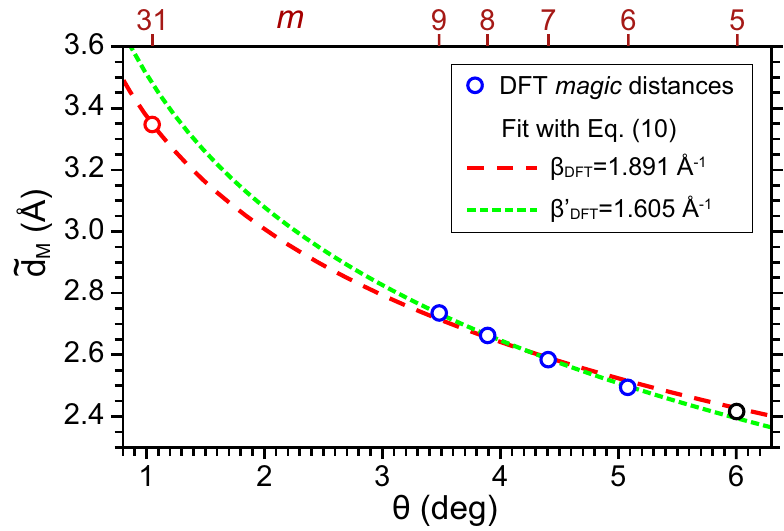}
\caption{(Blue circled points) DFT-derived magic distances $\tilde{d}_{\msc{ma}}{=}\tilde{d}_{m}$ \textit{vs.} commensurate twist angles $\theta{=}\theta_{m}$~(bottom  horizontal axis) for $m{=}6,{\ldots},9$~(top horizontal axis).
(Red circled point) MA condition established at the TB level: $d_{0}{=}3.349$~{\AA} and $\theta_{0}{=}1.05^{\circ}$.
(Red dashed line) Function in Eq.~(\ref{dma}), derived using the reference values $d_{0}$, $\theta_{0}$, and the best-fit parameter $\beta_{\msc{DFT}}{=}1.891$~{\AA}$^{-1}$.
(Green dotted line) Function in Eq.~(\ref{dma}), with parameters $\beta^{\prime}_{\msc{DFT}}{=}1.605$~{\AA}$^{-1}$ and, equivalently, $\theta^{\prime}_{0}{=}1.29^{\circ}$ at fixed $d_{0}$, or $d^{\prime}_{0}{=}3.480$~{\AA} at fixed $\theta_{0}$.
(Black circled point) Interlayer distance $\tilde{d}_{5}$ at the twisted angle $\theta_{5}$, excluded from the fitting processes above.\label{figdMdft}}
\end{figure}

As a final convincing evidence, we emphasize that the DFT-derived magic distances are well correlated with the commensurate twist angles through the same function introduced at the TB level, cf. Eq.~(\ref{dma}), which continues to serve as a reliable approximation of the bijective function~(\ref{dt}).
This identification is detailed in Fig~\ref{figdMdft}, where the optimal value $\beta_{\msc{DFT}}{=}1.891$~{\AA}$^{-1}$, for the inverse damping length, was obtained with the same reference parameters, $\theta_0{=}1.05^{\circ}$ and $d_0{=}3.349$~{\AA}, as in the TB approach.
The resulting fitting function~(red dashed line) slightly underestimates the $m{=}9$ point and slightly overestimates the $m{=}6$ point.
By contrast, optimizing both $\beta$ and $\theta_{0}$ with the interlayer distance fixed at $d_{0}{=}3.349$~{\AA} yielded best-fit parameters $\beta^{\prime}_{\msc{DFT}}{=}1.605$~{\AA}$^{-1}$ and $\theta^{\prime}_{0}{=}1.29^{\circ}$.
This adjustment significantly improved the agreement~(green dotted line), suggesting that the PBE-DFT approach predicts a barely larger MA at the nominal interlayer distance. 
Equivalently, optimizing $\beta$ and $d_0$ at the fixed MA $\theta_0$ yielded $\beta^{\prime}_{\msc{DFT}}{=}1.605$~{\AA}$^{-1}$ and $d^{\prime}_{0}{=}3.480$~{\AA}~(shown by the same green dotted line), which aligns with the known tendency of the PBE functional to underestimate interlayer coupling.

\section{Discussion\label{DsCsec}}
We have addressed the challenging question of whether the equivalence class for MATBG, firmly established within the CM framework, can be extended to TB models and, more importantly, to \textit{ab~initio} methods such as DFT.
Our results demonstrate that it is possible to define a magic condition linking the twist angle to the interlayer distance, for arbitrary twist angles within the range of validity of our approach ($\theta\lesssim5^{\circ}$). For both the specific TB approach and at the DFT-PBE level, we find that twist angle and interlayer distance under magic conditions are related by a simple function, Eq.~(\ref{dma}), which depends on a single parameter: the inverse decay length $\beta$. Given the simplicity of this relation and the consistent values of $\beta$ obtained in both approaches, we contend that our findings are robust and independent of the specific modeling employed. Consequently, the magic distances extracted from Figs.~\ref{figdma} and~\ref{figdMdft} provide reliable starting points for analogous search procedures, should one adopt alternative TB parametrizations or different exchange-correlation functionals and pseudopotentials within DFT.

The identification of the magic TBG set provides an effective strategy for studying ground-state and excited-state properties using geometrically simplified, computationally feasible structures.
As a prominent example, we suggest investigating the dynamical density-density response function of TBG electrons within the quasi-flat band region~\cite{Giuliani:826125}.
Using the equivalence-class approach, time dependent~(TD) DFT computations are accessible for these system and could offer valuable insights into plasmon excitations in MATBG. A full ab initio modelling of these properties could help understand the role of collective excitation in light trapping phenomena in TBG~\cite{stepanov2020untying,ge2021independent} or in the emergence of unconventional  superconductivity~\cite{PhysRevResearch.2.022040,peng2023theoretical,cea2021coulomb}.
Building upon this concept, the equivalence-class framework could be employed to study the superconducting-to-normal state transition in TBG systems with quasi-flat bands at specific magic distances.
Within this framework, the critical temperature is expected to be renormalized following a scaling relation analogous to that expressed in Eq.~(\ref{eqscale}).

Finally, we emphasize that the equivalence-relation concept is not limited to TBG.
It can be extended to other twisted multilayer materials comprising two or more atomic layers that exhibit Dirac-cone states in their isolated phases.
In these systems, the weak interlayer interaction should, under appropriate limiting conditions, yield a reference Hamiltonian analogous to the BM model, expressible in a form similar to Eq.~(\ref{MC}).
A particularly challenging but critical aspect, especially in heterogeneous multilayer structures, is the development of a smooth, bijective function that maps hopping parameters to interlayer distances.

In summary, we have demonstrated that MATBG belongs to a specific equivalence class, well-defined within the framework of the CM.
This equivalence class encompasses TBG structures characterized by a twist angle and a \textit{magic} interlayer distance.
Members of this class share homothetic electronic spectra, a fixed number of bands with vanishing Fermi velocity, and a common eigenstate basis that can be mapped one-to-one to those of MATBG via a dimensionless scaling factor associated with the twist angle.
The equivalence relation extends naturally to TB  models through a systematic search for the optimal interlayer distance at a fixed twist angle.
This process identifies the configuration that achieves the best homothetic match between the quasi-flat bands of a given TBG structure and those of MATBG.
Focusing on commensurate lattices, we established a one-to-one smooth correspondence between TBG structures with twist angles below  $6^\circ$ and MATBG.
Additionally, we investigated the ground-state electronic properties of twisted graphene bilayers within the framework of DFT, focusing on the same angular range.
An \textit{ab~initio} procedure was developed to determine the optimal interlayer distance for each twist angle that leads to the formation of quasi-flat bands.
Our findings open a pathway for employing \textit{ab~initio} methods to study systems with manageable atom counts per unit cell, while retaining the essential electronic properties of MATBG.

\section{Methods}
\subsection{TB computational details\label{TBdetails}}
A typical TB calculation scheme~\cite{PhysRevB.98.081410, PhysRevB.98.235137} was implemented, utilising $p_z$ orbitals centered at the atomic positions $\rv_{i}$ of the TBG lattice.
The hopping amplitudes $t_{ij}$ between distinct atomic sites $i$ and $j$ were described within the standard SK formalism~\cite{PhysRev.94.1498} as:
\begin{equation}
t_{ij}= (1-n_{ij}^2) V_{pp\pi}(r_{ij}) + n_{ij}^2 V_{pp\sigma}(r_{ij}).
\label{tij}
\end{equation}
Here, $V_{pp\sigma}$ and $V_{pp\pi}$ denote the interlayer and intralayer potentials, respectively, which depend on the inter-site distance $r_{ij}{=}|\rv_{j} {-}\rv_{i}|$.
The interplay between these interactions is modulated by the out-of-plane directional cosines $n_{ij}{=}z_{ij}/r_{ij}$, with $z_{ij}$ representing the perpendicular component of $\rv_{j}{-}\rv_{i}$ relative to the graphene sheets.

The potentials $V_{pp\sigma}$ and $V_{pp\pi}$ were modelled using a non-standard SK parameterization~\cite{PhysRevB.101.155405}, expressed as:
\begin{align}
V_{pp\sigma}(r_{ij})=&t_1 e^{-\eta(d-d_{0})} e^{-\delta(r_{ij}-d)} c(r_{ij}),
\label{Vpps}\\
V_{pp\pi}(r_{ij})=&-t_0 e^{-\delta(r_{ij}-a_{c})} c(r_{ij}),
\label{Vppp}
\end{align}
where the exponential factors account for in-plane structural deformation and out-of-plane atomic corrugation effects, controlled by $\eta$ and $\delta$, respectively, and
$c(r_{ij}){=}[1{+}e^{(r_{ij}-r_c)/l_c}]^{-1}$,
introduces a smooth cutoff to ensure convergence.
The parameter values adopted in this study were as follows: nearest-neighbor interatomic distance $a_{c}{=}1.42$~{\AA}, equilibrium interlayer distance without applied pressure $d_{0}{=}3.349$~{\AA}, intralayer hopping energy $t_{0}{=}2.8$~eV, interlayer hopping energy $t_{1}{=}0.44$~eV, inverse structural length $\delta{=}2.218$~{\AA}$^{-1}$, inverse corrugation length $\eta{=}0.58$~{\AA}$^{-1}$, smooth function characteristic distance $l_{c}{=}0.265$~{\AA}, and smooth function decay length $r_{c}{=}5.0$~{\AA}.
The calculations incorporated all hopping interactions within $r_{ij}{\lesssim}7.5${\AA}, implemented via the TBPLas package\cite{LI2023108632}.

To establish the MA condition, we focused on the commensurate twist angles $\theta_{m}$ defined in Eq.~(\ref{eq2}).
Initially, compression effects were neglected by setting $d{=}d_{0}$ in Eq.~(\ref{Vpps}), corresponding to the standard SK parameterization~\cite{PhysRev.94.1498}.
This approach facilitated the identification of the MA at $\theta_{31}{=}1.050^\circ$ and the computation of the MATBG eigenspectrum, along with its corresponding eigenstates.
Subsequently, the analysis was extended to a broad range of $\theta_{m}$ values greater than the MA, with $5{\leq}m{<}31$.
For each $\theta_{m}$, the interlayer distance $d$ was systematically varied to determine the corresponding magic distances $d_{m}$ through an iterative search procedure.

\subsection{DFT calculations\label{DFTdetails}}
DFT calculations were performed using the Quantum Espresso~(QE) package~\cite{Giannozzi_2009,Giannozzi_2017} within the projector augmented wave (PAW) method~\cite{PhysRevB.50.17953} using the PseudoDojo PAW datasets~\cite{Jollet2014}.
The PBE exhange-correlation functional~\cite{PBE_xc} was employed, with a plane wave energy cutoff of $30$~Ry and an energy convergence threshold of $10^{-7}$~Ry for the self-consistent runs.
A vacuum region of $20$~{\AA} was introduced along the out-of-plane direction to model the TBG slab.
The {\fbz} was sampled using a $12{\times}12{\times}1$ Monkhorst-Pack grid~\cite{MonkPack} in conjunction with a Marzari-Vanderbilt smearing function~\cite{Marzarivanderbilt} with a width of $0.005$~Ry.

The chosen TBG geometries corresponded to commensurate lattices with twist angles ranging from $\theta_{5}$ to $\theta_{9}$, as defined in Eq.~(\ref{eq2}).
Each structure was generated by starting from an AA-stacked bilayer graphene slab and applying a counter-clockwise rotation of the top layer by the specified twist angle $\theta$ about a vertical axis passing through a carbon atom. Additional positions for the rotation axis were also considered, as described in Supplementary Section~{IV}. 

A DFT-based search was then conducted to determine the magic distances $\tilde{d}_{5}$ to $\tilde{d}_{9}$, as discussed in the main text. 
This involved selecting the four bands closest to the Fermi level and calculating their BW over the $12{\times}12{\times}1$ $k$-point grid, as shown in Fig.~\ref{figBW}.

\backmatter

\section*{Additional information}
\textbf{Supplementary information}
Supplementary information for this article is available at \emph{article-url}.

\section*{Data Availability}
 The data that support the findings of this study are available from the corresponding author upon  request.

\section*{Code Availability}
The Tight Binding calculations have been performed using the freely distributed TBPlas package (https://www.tbplas.net/).
The Density Functioal Theory calculations have been performed using the freely distributed Quantum Espresso package (https://www.quantum-espresso.org/).
The details needed to reproduce the computations have been provided in the “Methods” section.

\section*{Acknowledgements}
\def\cinecaurl{http://www.cineca.it/}
\def\cinecainfn{http://www.hpc.cineca.it/news/framework-collaboration-agreement-signed-between-cineca-and-infn}
\def\infnprog{INF16\_npqcd}
{\noindent}We thank N. Lo Gullo, J. Settino, and F. Plastina for useful discussion.
This research was partially supported \emph{Centro Nazionale di Ricerca in High-Performance Computing, Big Data and Quantum Computing}, PNRR 4 2 1.4, CI CN00000013, CUP H23C22000360005.
We acknowledge the \emph{Marconi}, \emph{Marconi}100, \emph{Galileo}100 and \textit{Leonardo} high performance computing resources, provided by the \href{\cinecaurl}{CINECA consortium (Italy)}, within the {\infnprog} project, under the \href{\cinecainfn}{CINECA-INFN} agreement.
We also acknowledge the \emph{Newton} and \emph{Alarico} high performance computing clusters, provided by the University of Calabria.

\section*{Author contributions}
AP introduced the equivalent class approach, performed the TB calculations, performed part of the DFT simulations, and contributed to the writing of the manuscript.
MP developed the equivalent class approach, supervised the TB calculations, performed another part of the DFT simulations, and contributed to the writing of the manuscript. 
AS supervised the entire work, cross-checked the TB calculations, cross-checked all DFT simulations, and wrote the manuscript.
All authors contributed equally to the interpretation and understanding of results and data analysis.

\section*{Competing interests}
The authors declare no competing interests.




\clearpage

\end{document}